\catcode`\@=11  
\def\answ{t}    
\def\prepri{p }      
\def\twocol{t }      

\if\answ\twocol   
\documentstyle[aps,prd,floats,eqsecnum,twocolumn,epsf]{revtex}
\else
\documentstyle[aps,prd,floats,eqsecnum,epsf]{revtex}
\fi

\begin{document}

\draft


\if\answ\twocol
    \twocolumn[\hsize\textwidth\columnwidth\hsize\csname
              @twocolumnfalse\endcsname
\fi


\title{Charged Brane-World Black Holes}

\author{Andrew Chamblin
}
\address{
{\tt chamblin@ctpblack.mit.edu} \\
Center for Theoretical Physics, MIT, Bldg. 6-304, 77 Massachusetts Ave.,
Cambridge, MA 02139
}

\author{Harvey S. Reall
}
\address{
{\tt H.Reall@damtp.cam.ac.uk}\\
DAMTP, Centre for Mathematical Sciences, University of
Cambridge,\\ Wilberforce Road, Cambridge CB3 0WA, United Kingdom
}

\author{Hisa-aki Shinkai
}
\address{
{\tt shinkai@gravity.phys.psu.edu} \\
Centre for Gravitational Physics and Geometry,
104 Davey Lab., Department of Physics,\\
The Pennsylvania State University,
University Park, Pennsylvania 16802-6300
}
\author{Tetsuya Shiromizu
}
\address{
{\tt siromizu@utap.phys.s.u-tokyo.ac.jp} \\
MPI fur Gravitationsphysik, Albert-Einstein Institut, D-14476 Golm, Germany
\\
Department of Physics, The University of Tokyo, Tokyo 113-0033, Japan\\
and
Research Centre for the Early Universe(RESCEU),
The University of Tokyo, Tokyo 113-0033, Japan}

\date{November 2, 2000 (revised version) ~~ to appear in Phys.
Rev. D. ~~ hep-th/0008177}

\maketitle

\if\answ\twocol
    \widetext
\fi

\begin{abstract}
We study charged brane-world black holes in the model of Randall and
Sundrum in which our universe is viewed as a domain wall in asymptotically
anti-de Sitter space. Such black holes can carry two types of ``charge'',
one arising from the bulk Weyl tensor and one from a gauge field
trapped on the wall. We use a combination of analytical and numerical
techniques
to study how these black holes behave in the bulk. It has been shown that
a Reissner-Nordstrom geometry is induced on the wall when only Weyl charge
is present. However, we show that such solutions exhibit pathological
features in the bulk. For more general charged black holes, our
results suggest that the extent of the horizon in the fifth
dimension is usually less than for an uncharged black hole
that has the same mass or the same horizon radius on the wall.
\end{abstract}


\if\answ\twocol
    \vskip 2pc]
    \narrowtext
\fi

\section{Introduction}

In many of the brane-world scenarios, the matter
fields which we observe are trapped
on the brane\cite{add,aadd,RS1,RS2} (see also \cite{old} for
older proposals). If matter trapped on a brane undergoes
gravitational collapse then a
black hole will form. Such a black hole will have a horizon that
extends into the dimensions transverse to the brane: it will be a
higher dimensional object.

Within the context of the second Randall-Sundrum (RS) scenario\cite{RS2},
it is important that the induced metric on the domain wall\footnote{In
this paper, we use the terms ``domain wall'' and ``brane''
interchangeably.} is, to a good
approximation, the solution predicted by standard General Relativity
in four dimensions.  Otherwise the usual astrophysical properties
of black holes and stars would not be recovered.

In a recent paper \cite{Andrew}, the gravitational collapse
of {\it uncharged}, non-rotating matter
in the second model of RS was investigated.
There it was proposed that what would appear to be a four-dimensional black
hole from the point of view of an observer in the brane-world, is really
a five-dimensional ``black cigar'', which extends into the extra fifth
dimension.  If this cigar extends all the way down to the anti-de
Sitter (AdS) horizon,
then we recover the metric for a black string in AdS.  However,
such a black string is unstable near the AdS
horizon.  This instability, known as the ``Gregory-Laflamme'' instability
\cite{Ruth},
implies that the string will fragment in the region near
the AdS horizon.   However, the solution is stable far from the AdS
horizon.  Thus, one may conclude that the late time solution
describes an object that looks like the black string
far from the AdS horizon (so the metric on the domain wall is close to
Schwarzschild) but has a horizon that closes off before reaching the
AdS horizon. A similar effect occurs when there is more than one
extra dimension transverse to the brane \cite{csaba}.
These conclusions are supported by an exact calculation
for a three dimensional RS model \cite{Gary}.

In this paper, we consider black holes
charged under gauge fields which are {\it trapped on the brane}.
The flux lines of such gauge fields can pierce the horizon
only where it actually intersects the brane. The bulk
theory is the same as for the uncharged case so
one might expect that the black cigar solution would still
describe the bulk metric of such a charged brane-world black hole. The
effect of the charge might simply be to modify the position of the
brane in the bulk spacetime. If this were the case, then we might be
able to repeat the analysis of \cite{Andrew} by starting with the
black string metric and solving the Israel equations appropriate for
the presence of a gauge field on the brane.
However, in the Appendix we prove that this is not possible. It is
still conceivable that the bulk metric is the same as that of the
black {\it cigar}, but unfortunately the form of the cigar metric is not
known. We are therefore forced to study charged brane-world black holes
numerically.



A recent paper \cite{Dad} has claimed to give a solution describing a
non-charged black hole in the RS scenario. By using the
brane-world Einstein equations derived in \cite{Tess}, it was shown
that a Reissner-Nordstrom (RN) geometry could arise on the
domain wall provided that
the bulk Weyl tensor take a particular form at the wall.
We regard this solution as unsatisfactory for two reasons. First,
there is no Maxwell field on the domain wall so the black hole cannot
be regarded as charged\footnote{
In the AdS/CFT interpretation \cite{Gubser} of the RS model, this
black hole must be charged with respect to a $U(1)$ subgroup of the
dual CFT.}.
Secondly, only the induced metric on the
domain wall was given -- the bulk metric was not discussed. The solution
is simply a solution to the Hamiltonian constraint of general
relativity and gives appropriate initial data for evolution into the
bulk.
Until this evolution is performed and boundary conditions
in the bulk are imposed, it is not clear what this
solution represents. For example, it might give rise to some pathology
such as a naked curvature singularity. We would then not regard it as
a brane-world black hole, which should have a regular horizon
\cite{Andrew,Gary}. One aim of the present paper is to evolve the
initial data of \cite{Dad} in order to understand what this
``solution'' really describes.

The second aim of this paper is to study brane-world black holes that
are charged with respect to a Maxwell field on the brane. We start by
solving the Hamiltonian constraint on the brane to give an induced
metric that is close to, but not exactly, Reissner-Nordstrom. The
``charge'' of \cite{Dad} arises as an integration constant in the
metric. We then evolve this ``initial'' data away from the domain wall in
order to study the resulting bulk spacetime. Our solution to the
Hamiltonian constraint is based on a metric ansatz that is almost
certainly not obeyed by the true solution describing a charged
brane-world black hole. However, we expect our ansatz to be
sufficiently close to the true solution that useful results can be
obtained without a knowledge of the exact metric, just as in
\cite{Andrew}.

Our results suggest that it is more natural to take the ``charge
squared'' parameter of \cite{Dad} to be negative than positive since
the latter gives an apparent horizon that grows relative to the black
string as one moves away from the brane. For black holes charged with
respect to a Maxwell field, we find that the horizon shrinks in the
fifth dimension. In both cases (and for black holes carrying both
charges), we obtain a numerical upper bound on the length of the
horizon in the fifth dimension. We find that increasing either type of
charge tends to decrease this length, even if the horizon radius on the
brane is held fixed.

It is worth emphasizing that this paper is quite distinct from
recent papers which have appeared on the subject of charged black
holes in brane-world scenarios \cite{kaloper,lu,oda}.
This is because these papers all
study the effects of {\it bulk} charges on the brane-world geometry,
whereas our analysis deals with gauge degrees of freedom that
are truly localized on the brane. One consistent interpretation of the
RN solution of \cite{Dad} would be as the induced metric on the brane
in the (bulk) charged black string solution of \cite{lu,oda}. However,
in this paper we will study whether sense can be made of this solution
without introducing bulk gauge fields.

Related numerical work on uncharged brane-world black holes
has recently appeared in \cite{Shibata}. The
difference between that paper and the present work is that we will
prescribe ``initial'' data on the brane and evolve it in the spacelike
direction transverse to the brane, whereas in \cite{Shibata}, initial
data was prescribed on a spacelike hypersurface and evolved in a timelike
direction.

The outline of this paper is as follows.  First, we set up the basic
notation and formalism for a covariant treatment of the second brane-world
scenario of Randall and Sundrum. Next we solve the Hamiltonian
constraint for ``initial'' data on the brane and obtain a RN solution
with small corrections. We then numerically evolve the solution into the bulk
subject to the constraint that the metric solve the vacuum Einstein equations
with a negative cosmological constant. Finally we discuss the properties
of the resulting bulk spacetime.


\section{Formulation and strategy}

\subsection{Covariant formulation of brane-world gravity}

We shall be discussing a thin domain wall in a five dimensional bulk
spacetime. We shall assume that the spacetime is symmetric under
reflections in the wall. The 5-dimensional Einstein equation is
%
\begin{eqnarray}
{}^{(5)\!}R_{\mu\nu}
-\frac{1}{2}{}^{(5)\!}g_{\mu\nu}{}^{(5)\!}R=\kappa_5^2 \,
{}^{(5)\!}T_{\mu\nu},
\end{eqnarray}
%
where $\kappa_5^2 = 8\pi G_5$ and $G_5$ is the five dimensional Newton
constant. The energy-momentum tensor has the form
%
\begin{eqnarray}
{}^{(5)\!}T^{\mu\nu}=-\Lambda_5 {}^{(5)\!}g_{\mu\nu}+\delta (\chi)
[-\lambda h_{\mu\nu}+T_{\mu\nu}].
\end{eqnarray}
%
In the above, the brane is assumed to be located at $\chi=0$ where
$\chi$ is a Gaussian normal coordinate to the domain wall.
$\chi=0$ is the fixed point of the $Z_2$ reflection symmetry.
$\Lambda$ and
$\lambda$ denote the bulk cosmological constant and the domain wall
tension respectively. $h_{\mu\nu}$ is the induced metric on the
wall, given by $h_{\mu\nu} = {}^{(5)\!}g_{\mu\nu} - n_{\mu} n_{\nu}$
where $n_{\mu}$ is the unit normal to the wall.
The effect of the singular source at $\chi=0$ is described by
Israel's junction condition \cite{Israel}
%
\begin{eqnarray}
     K_{\mu\nu}|_{\chi =0}=-\frac{1}{6}\kappa_5^2 \lambda h_{\mu\nu}
     -\frac{1}{2}\kappa_5^2\Bigl(T_{\mu\nu}-\frac{1}{3}h_{\mu\nu}T \Bigr).
\label{eq:ext}
\end{eqnarray}
%
Here, $K_{\mu\nu}$ denotes the extrinsic curvature of the domain
wall, defined by $K_{\mu\nu} = h_{\mu}^{\rho} h_{\nu}^{\sigma}
\nabla_{\rho} n_{\sigma}$. In equation (\ref{eq:ext}), we are
calculating the extrinsic curvature on the side of the domain wall
that the normal point {\it into}. This is because we want to evolve
initial date prescribed on the wall in the direction of this normal.
Using the Gauss equation and the junction condition,
we recover the Einstein equation on the brane\cite{Tess}:
%
\begin{eqnarray}
\label{eqn:braneein}
{}^{(4)\!}G_{\mu\nu}=-\Lambda_4 h_{\mu\nu}
+8\pi G_4 T_{\mu\nu}+\kappa_5^4\pi_{\mu\nu}-E_{\mu\nu},
\end{eqnarray}
%
where
%
\begin{eqnarray}
\Lambda_4&=&\frac{1}{2}\kappa_5^2\Bigl(
\Lambda_5+\frac{1}{6}\kappa_5^2\lambda^2
\Bigr)\\
G_4&=&\frac{\kappa^4_5 \lambda}{48\pi}\\
\pi_{\mu\nu}&=&
\frac{1}{12}TT_{\mu\nu}
-\frac{1}{4}T_{\mu\alpha}T^\alpha_\nu
+\frac{1}{8}h_{\mu\nu}T_{\alpha\beta}T^{\alpha\beta}
-\frac{1}{24}h_{\mu\nu}T^2
\end{eqnarray}
%
and $E_{\mu\nu}$ is the `electric' part of the
5-dimensional Weyl tensor:
%
\begin{equation}
E_{\mu\nu} = {}^{(5)\!}C_{\mu\alpha\nu\beta} n^\alpha n^\beta
\end{equation}
%
We shall now specialize to the RS model. This has
\begin{equation}
     \Lambda_5 = -\frac{6}{\kappa_5^2 \ell^2}, \qquad \lambda =
     \frac{6}{\kappa_5^2 \ell},
\end{equation}
which implies
\begin{equation}
     \Lambda_4 = 0, \qquad G_4 = \frac{G_5}{\ell}. \label{impliedquantity}
\end{equation}
The matter on the domain wall will be assumed to be a Maxwell field.
This implies $T=0$, so we can rewrite the Einstein equation as
%
\begin{eqnarray}
{}^{(4)\!}R_{\mu\nu}=
8\pi G_4 T_{\mu\nu}-\frac{\kappa_5^4}{4}T_{\mu\rho}T_\nu^{\rho}
-E_{\mu\nu}. \label{eq:einstein2}
\end{eqnarray}
%
The Israel equation gives the extrinsic curvature of the wall:
%
\begin{eqnarray}
\label{eqn:israel}
K_{\mu\nu}|_{\chi=0}=-\frac{1}{\ell}h_{\mu\nu}
-\frac{\kappa_5^2}{2}T_{\mu\nu}.
\end{eqnarray}
%

\subsection{Strategy}

We adopt the following procedure: We take a certain charged black hole
geometry for the brane. When we solve for the bulk, we Wick rotate
twice. This gives a Kaluza-Klein bubble spacetime
\cite{KK1,KK2} from which we obtain boundary conditions at the
condition on the bubble surface. Wick rotating back gives boundary
conditions at the bulk horizon for our problem. The
Kaluza-Klein bubble spacetime is reviewed in Appendix B.

\subsection{Metric and field equations}

We assume that the induced metric on the brane takes the form
%
\begin{eqnarray}
ds^2 = -U(r)dt^2+\frac{dr^2}{U(r)}
+r^2 d\Omega_2^2,  \label{eq:metric}
\end{eqnarray}
%
where $d\Omega_2^2=d\theta^2 + \sin^2 \theta d \varphi^2$.
Note that this is a {\it guess}. It is unlikely that the exact metric
describing a brane-world black hole would have precisely this form
- in general one would expect the coefficients of $dt^2$ and $dr^2$ to be
independent (when the coefficient of $d\Omega_2^2$ is fixed as $r^2$).
However, we know that the induced metric describing a
charged black hole should be close to Reissner-Nordstrom, which {\it
can} be written in this form, so our ansatz is probably quite a good
guess. We expect that deviations from the exact metric will give rise to
pathologies when this initial data is evolved into the bulk. Even so,
the analysis of \cite{Andrew} shows that it is possible to extract
quite a lot of information from a pathological solution. The function
$U(r)$ will be determined from the Hamiltonian constraint equation below.
The bulk metric is assumed to take the form
%
\begin{eqnarray}
ds^2 &=& N(\chi,r)^2 d\chi^2-e^{2a(\chi,r)}U(r)dt^2 \nonumber \\ &&
+\frac{e^{2b(\chi,r)}dr^2}{U(r)}+e^{2c(\chi,r)}r^2 d\Omega_2^2,
\label{eq:metric2}
\end{eqnarray}
$N$ is the lapse function which describes the embedding
geometry of the hypersurface spanned by $(t, r, \theta, \varphi)$
during the evolution in the bulk spacetime.

The extrinsic curvature of a hypersurface of constant $\chi$ (with
unit normal $n = N d\chi$) is given by
\begin{eqnarray}
\label{eqn:extrinsiccpts}
K^t_t=\frac{{\dot a}}{N},\quad K^r_r
=\frac{{\dot b}}{N} \quad {\rm and} \quad
K^\theta_\theta=K^\varphi_\varphi
= \frac{{\dot c}}{N}, \label{KIJs}
\end{eqnarray}
where a dot denotes $\partial_\chi$.
The spacetime is described by the evolution equation,
\begin{eqnarray}
{\dot K}^\mu_\nu&=&N\Bigl({}^{(4)\!}R^\mu_\nu-KK^\mu_\nu
+\frac{4}{\ell^2}\delta^\mu_\nu \Bigr)-D^\mu D_\nu N,  \label{Kdot}
\end{eqnarray}
the Hamiltonian constraint equation,
%
%
\begin{eqnarray}
{}^{(4)\!}R-K^2+K_{\mu\nu}K^{\mu\nu}=-\frac{12}{\ell^2},
\label{eq:Hami}
\end{eqnarray}
%
and the momentum constraint equation,
%
\begin{eqnarray}
D_\mu K^\mu_\nu-D_\nu K=0. \label{eq:CM}
\end{eqnarray}
%
Here ${}^{(4)\!}R^\mu_\nu$ and ${}^{(4)\!}R$ are
the Ricci tensor and Ricci scalar on hypersurfaces of constant $\chi$.

\section{Brane and Bulk Geometry}
\subsection{Brane Geometry : Charged black hole ``initial data"}

The action for the Maxwell field on the brane is taken to be
\begin{equation}
     S = -\frac{1}{16 \pi G_4} \int d^4 x \sqrt{-h} F_{\mu\nu} F^{\mu\nu},
\end{equation}
giving energy-momentum tensor
\begin{equation}
     T_{\mu\nu} = \frac{1}{4\pi G_4} \left(F_{\mu \rho} F_{\nu}\,^{\rho} -
     \frac{1}{4} h_{\mu\nu} F_{\rho\sigma} F^{\rho\sigma} \right).
\end{equation}
The field strength $F$ is related to a potential $A$ by $F = dA$. The
equations of motion are satisfied if we take $A = -\Phi(r) dt$ with
$\Phi(r) = Q/r$. This gives
\begin{equation}
     T_{\mu\nu} = \frac{1}{8\pi G_4} \frac{Q^2}{r^4} \mathrm{diag}
     \left(U, -U^{-1}, r^2, r^2 \sin^2 \theta \right).
\end{equation}
This can be substituted into the right hand side of the Israel
equation (\ref{eqn:israel}) to give an expression for the extrinsic
curvature. This can then be substituted into the Hamiltonian
equation (\ref{eq:Hami}),
along with our metric ansatz to give an equation for $U(r)$. Solving
this equation gives\footnote{
It is interesting to compare this form for $U(r)$ with the behaviour
expected from the linear perturbation analysis of the second
RS model\cite{RS2,Tama,Misao,Lisa}. In linearized theory, $U(r) = 1 -
\phi(r)$ where $\phi(r)$ is the Newtonian potential.
For $r \gg \ell$, the leading order corrections to $\phi(r)$
are expected to be proportional to $G_4 M \ell^2/r^3$ and $\ell^2 Q^2 / r^4$.
Such terms are not present in our expression for $U(r)$. However, we
shall be interested in black holes for which $G_4 M \gg \ell$, so
these correction terms will be small compared with terms like $(G_4
M/r)^3$ and $(G_4 M Q/r^2)^2$, which would be neglected in linearized
theory. In other words, the RS correction are dominated by
post-Newtonian corrections \cite{Lisa} so it is not appropriate to
compare $U(r)$ with the linearized results beyond leading order.}
\begin{equation}
     U(r) = 1 - \frac{2G_4 M}{r} + \frac{\beta + Q^2}{r^2}
+ \frac{l^2 Q^4}{20 r^6},
\end{equation}
where $M$ and $\beta$ are arbitrary constants of integration.
Substituting
into the Einstein equation on the domain wall gives
\begin{equation}
     -E_{\mu\nu} = \left( \frac{\beta}{r^4} + \frac{l^2 Q^4}{2 r^8}
     \right) \mathrm{diag} \left( U, -U^{-1}, r^2, r^2 \sin^2 \theta
     \right).
\end{equation}
It is interesting to compare $-E_{\mu\nu}$ with $8\pi G_4 T_{\mu\nu}$
since these quantities appear on an equal footing in the effective
Einstein equation (\ref{eqn:braneein}). It is
clear that the constant of integration $\beta$ is in some sense
analogous to $Q^2$, which is why the authors of \cite{Dad} obtained a
RN solution. However, since their solution did not
have a Maxwell field, it cannot really be regarded as a charged black
hole in the usual sense. Rather it carries ``tidal'' charge associated
with the bulk Weyl tensor. $\beta$ might be regarded as a {\it five}
dimensional mass parameter.

We shall only consider initial data that corresponds to an object with an
event horizon (in the four dimensional sense) on the domain wall. In some
cases there may be more than one horizon. We shall use $r_+$ to denote the
position of the outermost horizon, i.e., the largest solution of
$U(r)=0$. This has to be found numerically except when $Q=0$.

Our ``initial data" is given by
\begin{eqnarray}
\left. \frac{{\dot a}}{N} \right|_{\chi=0} &=&
\left. \frac{{\dot b}}{N} \right|_{\chi=0} =-\frac{1}{\ell}
+\frac{\ell}{2}\frac{Q^2}{r^4},
\\
\left. \frac{{\dot c}}{N}
\right|_{\chi=0} &=&
-\frac{1}{\ell}-\frac{\ell}{2}\frac{Q^2}{r^4},
\label{extrinsic_initial}
\\
a(\chi=0,r)&=&b(0,r)=c(0,r)=0.
\end{eqnarray}

We shall study the following cases:

\noindent
{\bf (i) No electromagnetic charge}, i.e., $Q = 0$. In this case, the
induced metric on the domain wall is exactly RN\cite{Dad}. The horizon
radius is
\begin{equation}
\label{eqn:rplus}
    r_+ = M + \sqrt{M^2-\beta}.
\end{equation}
The induced metric has a regular horizon if $\beta \le M^2$.
Note that there is nothing to stop us choosing
$\beta$ to be {\it negative}, which emphasizes the difference
between the solution
of \cite{Dad} and a charged black hole. If we take $\beta$ to be
negative then the induced metric has only one horizon, instead of the
two horizons of a non-extreme RN black hole.

\noindent
{\bf (ii) No tidal charge}, i.e., $\beta=0$. In this case, the induced
metric on the domain wall is Reissner-Nordstrom with a correction
term. Note that $-E_{\mu\nu}$ is non-zero but is of order $1/r^8$,
which suggests that the total ``tidal energy'' on the wall is zero.

We shall also consider the general case {\bf (iii) Both charges non-zero},
i.e., $\beta \ne 0$, $Q \ne 0$.

\subsection{Bulk Geometry}

The bulk geometry is obtained by integrating
(\ref{KIJs}) and (\ref{Kdot}) in the $\chi$-direction numerically.
We use the standard `free-evolution' method, that is we do not
solve the constraint equations (\ref{eq:Hami}) and (\ref{eq:CM}) during
the evolution, but instead use them to monitor the accuracy of the
simulation.

We obtain the solution numerically in the region $r_+ < r < r_e$ with $r_e
\sim 5 r_+$. Boundary conditions at $r=r_+$ are specified by first Wick
rotating $\chi = iT$, $t=i\tau$, which takes the metric to a Kaluza-Klein
bubble metric (see Appendix B).
Therefore we can apply the numerical techniques that are used in the
study of  Kaluza-Klein bubbles \cite{Shinkai},
although the physics of Kaluza-Klein bubbles is unrelated to the physics
of black holes.
It was shown in the Appendix of \cite{Shinkai} that
at the inner boundary $r=r_+$, $a$ and $b$ evolve synchronously, that is,
$a(T,r_+) = b(T,r_+)$. Analytically continuing back to our original
spacetime yields the boundary condition $a(\chi,r_+) = b(\chi,r_+)$. The
evolution equation for the trace of $K_{\mu\nu}$ and the momentum
constraint are also used at $r=r_+$.
At the outer boundary $r=r_e$, we assume the components of the extrinsic
curvature [equation (\ref{eqn:extrinsiccpts})] fall off
like $-1/\ell+ {\cal O}\left( r^{-4} \right)$ [cf.
(\ref{extrinsic_initial})].
We apply the geodesic gauge condition (slicing condition), $N=1$.

We use the Crank-Nicholson integrating scheme with two iterations
\cite{TeukolskyCrankNicholson}. The numerical
code passed convergence tests, and the results shown in this
paper are all obtained to acceptable accuracy.

We were only able to solve numerically in a region near the domain wall
with a maximum value for $\chi$ of ${\cal O}(1)$. This is because
the volume element of surfaces of constant $\chi$ decreases exponentially
as one moves away from the wall, just as in pure AdS. The evolution was
stopped when $\sqrt{-g}$ became too small to monitor accurately.

We are interested in how charge affects the shape of the
horizon, in particular how far it extends into the fifth dimension. This
will be measured by the ratio of the physical size of the apparent
horizon $r_+ e^{c(\chi,r_+)}$, to that of a black string \cite{Andrew}
with the same horizon radius $r_+$ on the wall\footnote{
The reason for
measuring the size of the horizon relative to that of the black string
is because we want to distinguish the closing-off of the horizon from
the exponential collapse of hypersurfaces of constant $\chi$ arising from
the AdS nature of the geometry.}.
The size of the black
string apparent horizon in the bulk is $r_+ e^{-\chi/\ell}$, so the ratio
is
\begin{eqnarray}
R(\chi)=e^{c(\chi,r_+)+\chi/\ell}. \label{AHstringratio}
\end{eqnarray}
We remark that the only apparent horizon that appears during the
$\chi$-evolution is at $r=r_+$. Here we define apparent horizon
as the outermost region of negative
expansion of the outgoing null geodesic congruences, where we
define the expansion rate, $\theta_+$, as
\begin{eqnarray}
\theta_+ &=& {}^{\! (3)}\nabla_a s^a
+ {}^{\! (3)}K - s^a s^b  {~}^{\! (3)}K_{ab},
    \label{expansion}
\end{eqnarray}
where $s^a=(1/\sqrt{g_{rr}}) \partial_r$ is an outwards pointing unit vector
in the $3$-dimensional metric.  We checked (\ref{expansion}) during
the evolution and confirmed its positivity for $r>r_+$.

%
{}Our initial conditions give the behaviour of
the ratio $R(\chi)$ near the brane:
%
\begin{eqnarray}
\dot R(\chi)|_{\chi=0}=-\frac{\ell}{2}\frac{Q^2}{r_+^4} \leq 0,
\end{eqnarray}
%
and
%
\begin{eqnarray}
\ddot R(\chi)|_{\chi=0}=\frac{3Q^2+\beta}{r_+^4}-\frac{\ell^2Q^4}{2r_+^8}.
\end{eqnarray}
%
For model (i) ($Q=0$),
$\dot R(\chi)|_{\chi=0}=0$, but $\ddot R(\chi)|_{\chi=0}=
{\beta}  / {r_+^4}$.
This gives
$\ddot R(\chi)|_{\chi=0}<0$ for the case with $\beta<0$,
which indicates that the ratio decreases, while
$\ddot R(\chi)|_{\chi=0}>0$ for the case with $\beta>0$,
which indicates that the ratio increases.
We have plotted the numerical results for this ratio
in Fig.\ref{fig1} (a) and (b) (henceforth we shall set $\ell = G_4 = 1$ and
assume $M \gg
1$, as appropriate for an astrophysical black hole.).
Fig.\ref{fig1} (a) and (b) suggests that
a negative value for $\beta$ is the natural choice since the apparent
horizon grows (relative to the black string) in the fifth dimension
when $\beta$ is positive.

For model (ii) ($\beta = 0$),
$\dot R(\chi)|_{\chi=0}<0$ and the ratio
always  decreases [see Fig.\ref{fig1} (c)].
Model (iii) ($Q\neq 0$ and $\beta\neq 0$) is non-trivial.
We present numerical results
in Fig.\ref{fig2}.  The plot is for
$M=5$, $Q=3$ and $\beta=0, \pm 5, \pm 10, \pm 15$,
where $\beta=15$ is close to the extreme\footnote{
By extreme, we mean that $U(r)$ has a double root at $r=r_+$.}
case for this choice of $Q$.
The qualitative features are combinations of the plots in Fig.\ref{fig1}.
Note that $\beta$ seems to have the greatest effect on the bulk evolution.
Again, the case with negative $\beta$ appears to be the natural choice
since positive $\beta$ gives a growing horizon.

\subsection{Bulk Geometry: extent of the horizon}

In this section, we shall estimate how far the horizon extends into the
fifth dimension by combining analytical and numerical work.
Following the
conjugate points theorem\cite{HawkingEllis}, we shall show that for a
charged black hole, the trace of the extrinsic curvature diverges
at a finite distance from the brane.
The trace of the evolutional equation is given by
%
\begin{eqnarray}
\dot K = {}^{(4)}R-K^2+\frac{16}{\ell^2}
= -K_{\mu\nu}K^{\mu\nu}+\frac{4}{\ell^2},
    \label{eq:trace}
\end{eqnarray}
%
where we used the Hamiltonian constraint in
the second line.
Now define $k_{\mu\nu}$ as
%
\begin{eqnarray}
K^\mu_\nu=:-\frac{1}{\ell}h^\mu_\nu+k^\mu_\nu.
\end{eqnarray}
%
The trace part of $k_{\mu\nu}$, $k=k^\mu_\mu$, is expected to measure the
volume expansion relative to the AdS ``background'' geometry.
In term of $k_{\mu\nu}$,  (\ref{eq:trace}) can be written as
%
\begin{eqnarray}
\dot k -\frac{2}{\ell}k+\frac{1}{4}k^2 =
-\tilde k_{\mu\nu} \tilde k^{\mu\nu} \leq 0, \label{eq:tracenew}
\end{eqnarray}
%
where $\tilde k_{\mu\nu}$ is the traceless part of $k_{\mu\nu}$.
On the brane the ``initial'' condition is
%
\begin{eqnarray}
k_{\mu\nu}|_{\rm brane}= \tilde k_{\mu\nu}|_{\rm brane}= -4\pi G_5
T_{\mu\nu},
\end{eqnarray}
%
which implies
%
\begin{eqnarray}
k|_{\rm brane}=0
\end{eqnarray}
%

For the case with $Q \neq 0$,
%
\begin{eqnarray}
\tilde k_{\mu\nu} \tilde k^{\mu\nu} >  0,
\end{eqnarray}
%
so
%
\begin{eqnarray}
{\dot k}|_{\rm brane} < 0
\end{eqnarray}
%
This implies that there is a $\chi=\chi_0$ such that
%
\begin{eqnarray}
k=k_0<0
\end{eqnarray}
%
{}From  (\ref{eq:tracenew}), one obtains
\begin{equation}
1 + \frac{8}{\ell |k|} \le \left(1 + \frac{8}{\ell |k_0|} \right)
e^{2(\chi_0 - \chi)/\ell},
\end{equation}
from which it follows that $k$ diverges before $\chi=\chi_{crit}$, where
%
\begin{eqnarray}
\chi_{crit}
= \chi_0+\frac{\ell}{2}{\rm log}\Bigl(1+ \frac{8}{\ell |k_0|}
\Bigr).
\label{eq-Xcrit}
\end{eqnarray}
%
The divergence in $k$ implies that $K$ also diverges. Near $\chi =
\chi_{crit}$, $|k|$ behaves like
\begin{equation}
\label{eqn:kdiv}
k \le -\frac{4}{\chi_{crit}-\chi}.
\end{equation}
The case with $Q=0$ is more difficult to analyze because ${\dot k}|_{\rm
brane} =0$. We can use equation (\ref{eqn:extrinsiccpts}) (with $N=1$)
to give
\begin{equation}
\label{eqn:kexpr}
   k |_{r=r_+} = 2 \partial_{\chi} \left( a + c + \frac{2\chi}{\ell}
   \right),
\end{equation}
where we have used the synchronous evolution boundary condition
$a=b$ at $r=r_+$. In Fig.\ref{fig3},
we have plotted $a+c+2\chi/\ell$ at $r=r_+$. It is clear from this
plot that $k$ becomes negative in the bulk when $\beta < 0$. In fact
$k$ also becomes negative when $\beta > 0$.
Thus, even in the $Q=0$
case, there exists a $\chi = \chi_0$ such that $k=k_0 < 0$. The
above argument can then be used to show that when $Q=0$ and $\beta \ne
0$, $K$ must diverge before
$\chi = \chi_{crit}$, where $\chi_{crit}$ is given by equation
(\ref{eq-Xcrit}).
We have therefore proved that if $Q \ne 0$ or $\beta \ne 0$ then $K$
diverges before $\chi = \chi_{crit}$.

It follows from equations  (\ref{eqn:kdiv}) and (\ref{eqn:kexpr}) that
\begin{equation}
   (a + c)|_{r=r_+} \le 2 \log \left( \chi_{crit} - \chi \right),
\end{equation}
which implies that $\sqrt{-g}$ tends to zero at least as fast as
$\left( \chi_{crit} - \chi \right)^4$ as $\chi \rightarrow
\chi_{crit}$.

Conservatively, the divergence of $K$ indicates that the geodesic slicing
has broken down (when $N=1$, $\partial_\chi$ is the tangent vector of
spacelike geodesics), in other words a caustic has
occurred. The numerical study therefore cannot be extended
further using this slicing.
This has, however, a physical meaning because the apparent horizon is
located at constant $r=r_+$ in the bulk. The horizon will encounter
the caustic before reaching the AdS Cauchy horizon. The caustic can
therefore be viewed as the endpoint of the horizon, i.e., the tip of
the black cigar. Our analysis has only shown that the geodesic slicing
must break down at the caustic so, in principle, this point may be
regular, i.e., there may exist a coordinate chart that covers a
neighbourhood of this point\footnote{It is not even clear from our
analysis whether the caustic occurs at a single point or is spread
over a region of spacetime.}. However, we do not regard this as very
likely. Our guess for the induced metric on the domain wall is
unlikely to be exactly correct, so in general we would expect some
pathology such as a naked curvature singularity to appear in the
bulk. We cannot check whether curvature invariants diverge at $\chi =
\chi_{crit}$ since our numerical evolution cannot be extended as far as
$\chi = \chi_{crit}$.

Whether the bulk solution is regular or not, equation (\ref{eq-Xcrit})
gives us an upper bound on the extent of the horizon in the direction
transverse to the domain wall, i.e., the length of the black cigar. We have
plotted this upper bound in Fig.\ref{fig4} taking the values for
$\chi_0$ and $k_0$ at the endpoint of our numerical evolution.
The first graph shows
how $\chi_{crit}$ depends on $Q$ and $\beta$ when $M$ is
fixed. Note that when $Q=\beta=0$, the numerical solution is simply
the black string\footnote{The reader may find this surprising since
the black string is unstable \cite{Andrew}, and small numerical errors
might be expected to act as perturbations. However, the string is
unstable to {\it long} wavelength perturbations, and the numerical
errors will only be relevant at short wavelengths.}, which has
$\chi_{crit} = \infty$. Increasing $Q$ clearly has the effect of decreasing
$\chi_{crit}$, which is not surprising since increasing $Q$ also
shrinks the horizon radius on the domain wall $r_+$. Perhaps more
surprising is that making $\beta$ more negative also appears to
decrease $\chi_{crit}$ even though this {\it increases} the horizon
radius on the wall [see equation (\ref{eqn:rplus})]. The solid curve on
this diagram has both $M$ and $r_+$ fixed. It is clear that
$\chi_{crit}$ decreases along this curve as $Q$ or
$\beta$ increases.

The second graph of Fig.\ref{fig4} plots the same curve (fixed $M$
and fixed $r_+$) for different values of $M$. The trend seems to be
the same in each case.

The final graph of Fig.\ref{fig4} is for fixed $r_+$ (rather than
fixed $M$). Increasing $\beta$ appears to decrease $\chi_{crit}$
when $Q$ is small but has no significant effect when $Q$ is large.
When $\beta$ is non-zero, increasing $Q$ has the effect of initially
slightly increasing $\chi_{crit}$, but ultimately decreases
it substantially. The gross trend appears to be that increasing either
type of charge leads to a decrease in the length of the horizon.

In most of these graphs, $\chi_{crit} < r_+$, so the extent of the
horizon in the fifth dimension is smaller than the horizon radius on
the domain wall, just as for the uncharged black cigar.


\section{Summary and Discussion}

In this paper we have studied charged black holes in the second RS
model. We have seen that two types of charge can arise on the brane,
one coming from the bulk Weyl tensor \cite{Dad} and one from a Maxwell
field {\it trapped on the brane}. Starting from an ansatz for the induced
metric on the brane, we have solved the constraint equations of 4+1
dimensional gravity to find metrics describing charged brane-world
black holes. In the absence of Maxwell charge,
one can obtain a Reissner-Nordstrom solution \cite{Dad}. If Maxwell
charge is included then one can obtain a geometry that is
Reissner-Nordstrom with small corrections.

Using these induced metrics as ``initial'' data, we have solved the bulk
field equations numerically. We have found that the RN solution of
\cite{Dad} has an apparent horizon that grows (relative to the black
string apparent horizon) in the dimension
transverse to the brane unless the ``charge squared'' parameter
$\beta$ is taken to be negative\footnote{
In \cite{Shinkai}, the evolution of Kaluza-Klein bubbles was studied
numerically and it was found that even though negative mass bubbles
start off with accelerating expansion \cite{Ted}, the acceleration
ultimately becomes negative. It is conceivable that
something analogous could happen here but we have found no evidence
for such behaviour.}.
It therefore seems unlikely that this solution
really corresponds to a charged brane-world black hole. Of course, if a
bulk gauge field is included then the work of \cite{Dad} (with $\beta
>0$) has a natural interpretation as the induced metric on the brane
arising from the charged black string solution of \cite{lu,oda}.

If $\beta < 0$ and/or $Q \ne 0$ then the horizon shrinks
relative to the black string horizon.
For all cases (including $\beta > 0$), we have found that the trace of the
extrinsic
curvature diverges at a finite distance from the brane, with
the volume element of the spacetime tending to zero. For $\beta \le
0$, we have interpreted this as the end point of the horizon of the black
hole. Our results suggest that increasing the charges of a brane-world
black hole will decrease the length of its horizon in the fifth
dimension, even when the horizon radius on the brane is kept fixed.
This implies that, by adjusting $Q$, one can change the five
dimensional horizon area while keeping the four dimensional horizon
area fixed. One might think that this would lead to a difference
between the entropies calculated from these horizon areas, which would
be bad news for hopes of recovering General Relativity as the
effective four dimensional theory of
gravity on the brane. However, the exponential decrease in the volume
element as one moves away from the brane implies that the dominant
contribution to the five dimensional area comes from the region of the
horizon that is closest to the brane \cite{Gary}. Changes near the
other end of the horizon give only subleading corrections to the five
dimensional area, allowing the four and five dimensional entropies to
agree at leading order.

We suspect that our solutions will generically have a curvature singularity at
the point where the trace of the extrinsic curvature diverges. This is
because it seems rather improbable that our ansatz for the induced
metric on the brane should turn out to be exactly right.
However, we expect that for each value of $Q$ there will be some value
of $\beta$ for which a small change in our initial data
would smooth out this singularity, leading to a regular geometry
describing a brane-world black hole carrying Maxwell charge $Q$.
This smoothing would probably not significantly affect the position
of the ``tip'' of the horizon,
for which we have obtained an upper bound on the distance
from the brane. This is to be contrasted with the uncharged case in
which one takes the induced metric on the brane to be Schwarzschild.
Evolving this into the bulk gives the black string metric, for which
the singularity occurs at the AdS horizon, which is at {\it infinite} proper
distance from the brane along spacelike geodesics. A small
perturbation of the metric on the brane takes one from the black
string to the black cigar, which has a regular AdS horizon and a black
hole horizon with a tip at finite distance from the brane.

For the black string, the stability analysis of \cite{Andrew} shows
that the horizon extends a distance of order $d = \ell \log (G_4
M/\ell)$ into the fifth dimension, so $d \ll r_+$. Our results
give only an upper bound for $d$ in the charged case. It would be nice
if the stability analysis could be extended to the charged
case. However, the instability only sets in when the proper radius of the
horizon becomes smaller than the anti-de Sitter length scale and we
were not able to extend our numerical evolution this far. Our upper bound
seems rather on the large side, since it appears to give $d \sim r_+$
for small $Q$ and $\beta$. However, for large $Q$ and/or $\beta$, 
figure \ref{fig4}(c) shows that $d \ll r_+$, so our upper bound is
probably tighter in this case.

The main outstanding problem remains to find the exact bulk metric
that describes a brane-world black hole. This was solved for uncharged
black holes in the 3 dimensional RS model by using the 4
dimensional AdS C-metric in the bulk \cite{Gary}. Unfortunately, the
higher dimensional generalization of this metric is not known.
It would be interesting to see whether charged black holes in the 3
dimensional RS model could be constructed by using the same bulk as in
\cite{Gary} but simply slicing along a different hypersurface. It
would also be interesting to use the methods of \cite{Tama,Misao,Lisa} to
find linearized solutions describing spherical distributions of matter
charged with respect to a brane-world gauge field.

\section*{Acknowledgments}
AC thanks Dan Freedman, Andreas Karch, Philip Mannheim,
Joe Minahan and Lisa Randall
for useful conversations. AC is partially supported by the U.S. Dept.
of Energy under cooperative research agreement DE-FC02-94ER40818.
HSR thanks Stephen Hawking for useful conversations.
AC and HSR thank the organizers of the
Santa Fe 2000 Summer Workshop on ``Supersymmetry, Branes and
Extra Dimensions'' for hospitality while this work was being completed.
HS appreciates the hospitality of the CGPG group, and was supported in part by 
NSF grants PHY-9800973, and the Everly research funds of Penn State. 
HS was supported by the Japan Society for the Promotion of Science
as a research fellow abroad. TS thanks D. Langlois for discussions.
His work is partially supported by the Yamada foundation.
Numerical computations were performed using machines at CGPG.

\appendix
\section{Brane bending and the black string}

One candidate for a black hole formed by gravitational collapse
of charged brane-world matter
on a domain wall in AdS is the black string solution in AdS, which
has the metric
\begin{equation}
     ds^2 = \frac{{\ell}^2}{z^2}(-U(r)dt^2 + U(r)^{-1}dr^2 +
r^2 d\Omega_2^2 + dz^2)
\end{equation}
where $U(r) = 1-2 G_4 M/r$.  As discussed in \cite{Andrew}, surfaces
of constant $z$ trivially satisfy the Israel matching conditions
provided that the tension
satisfies $\lambda = \pm {6} / {{{\kappa}_{5}}^2 {\ell}}$.  Thus, we
may slice the spacetime along such a surface of constant $z$, and match
to a mirror image, in order to obtain the Schwarzschild solution on the
domain wall.

We now want to consider what happens when we allow the black hole
to be electrically charged with respect to some $U(1)$ gauge field living
on the brane.  Thus, we must add in an extra term to the brane-world
stress energy tensor of the form
\begin{equation}
T_{\mu \nu} = \frac{1}{4\pi G_4}(F_{\mu \rho}F_{\nu}^{\rho} -
\frac{1}{4}q_{\mu \nu}F_{\rho \sigma}F^{\rho \sigma})
\end{equation}

\noindent where the electric gauge potential has the form
\begin{equation}
A = -{\Phi}(r)dt
\end{equation}

\noindent so that
\begin{equation}
F = {\Phi}^{\prime}(r)dt ~{\wedge}~ dr
\end{equation}

\noindent where $^{\prime}$ denotes differentiation with respect to
$r$.

Now, as a first guess we might try to support this stress-energy on
the brane by allowing the brane to bend in the black string background
in such a way that the extrinsic curvatures would still satisfy the
Israel equations.

In other words, we allow the position
$z$ of the brane to depend on the radial direction $r$.
Solving the Maxwell equations yields
\begin{equation}
   \Phi'(r) = -\frac{Q}{r^2} \left(1 + {z'}^2 U \right)^{1/2}
\end{equation}
To compute the
extrinsic curvature of the timelike hypersurface swept out by $z = z(r)$,
we introduce an orthonormal basis which consists of a unit normal vector
\begin{equation}
     n = \frac{\epsilon \ell}{z \sqrt{1+U{z'}^2}} \left(dz - z'
dr\right),
\end{equation}
where $\epsilon = \pm 1$, a unit timelike tangent
\begin{equation}
     u = \frac{z}{\ell}U^{-1/2} \frac{\partial}{\partial t},
\end{equation}
and the spacelike tangents
\begin{equation}
     t = \frac{z}{\ell}\sqrt{\frac{U}{1+U{z'}^2}} \left(z'
\frac{\partial}{\partial z} + \frac{\partial}{\partial r}\right),
\end{equation}
\begin{equation}
     e_{\phi} = \frac{z}{\ell r \sin \theta} \frac{\partial}{\partial \phi},
\end{equation}
\begin{equation}
     e_{\theta} = \frac{z}{\ell r} \frac{\partial}{\partial \theta}
\end{equation}
It follows that the non-vanishing components of the extrinsic
curvature in this basis are
\begin{equation}
     K_{uu} = \frac{\epsilon}{\ell \sqrt{1+U{z'}^2}}(1 + \frac{1}{2}U'zz'),
\end{equation}
\begin{equation}
     K_{\theta\theta} = K_{\phi\phi} = \frac{- \epsilon}{\ell \sqrt{1+U
{z'}^2}} \left(1 + \frac{U}{r}zz'\right),
\end{equation}
\begin{equation}
     K_{tt} = -\frac{\epsilon U}{\ell \left(1+U{z'}^2\right)^{3/2}} \left(
zz'' + {z'}^2 + U^{-1} + \frac{U'zz'}{2U}\right).
\end{equation}
Under the assumption of $Z_2$ symmetry, the Israel equations reduce to
(\ref{eqn:israel}). The three independent components of $K_{\mu\nu}$
give three independent equations:
\begin{eqnarray}
K_{tt} &=& \frac{1}{\ell} - \frac{z^4 Q^2}{2 \ell^3 r^4} \nonumber \\
K_{uu} &=& -\frac{1}{\ell} + \frac{z^4 Q^2}{2 \ell^3 r^4} \\
K_{\theta\theta} &=& \frac{1}{\ell} + \frac{z^4 Q^2}{2 \ell^3 r^4}. \nonumber
\end{eqnarray}
It is straightforward to show that it is impossible to
solve these three equations simultaneously unless one takes $Q=0$ and
$z=$constant, which is the uncharged solution of \cite{Andrew}.
It is therefore not possible
to support the stress-energy of a point charge by simply allowing the brane
to bend in the black string background.
It follows that the {\it bulk} has to change once the
brane-world charge is introduced.  In other words, brane-world charge
will induce changes in the bulk Weyl tensor, and this is exactly what
we have found in our numerical analysis.

\section{Kaluza-Klein bubble}

The double Wick rotation($\chi \to i t, t \to i \tau$) of the
metric of Eq. (\ref{eq:metric}) gives us the Euclidean induced metric:
%
\begin{eqnarray}
ds^2=U(r)d\tau^2+\frac{dr^2}{U(r)}+r^2 d\Omega_2^2.
\end{eqnarray}
%
The largest $r=r_+$ such that $U(r_+)=0$ is interpreted as the
position of the bubble surface. Around $r=r_+$, the
metric can be expanded
%
\begin{eqnarray}
ds^2 \simeq U'(r_+)(r-r_+)d\tau^2+\frac{dr^2}{U'(r_+)(r-r_+)}+r_+^2
d\Omega_2^2.
\end{eqnarray}
%
In term of the new coordinate $R:={\sqrt {r-r_+}}$,
%
\begin{eqnarray}
ds^2 \simeq \frac{4}{U'(r_+)}\Bigl[R^2d \Bigl(
\frac{U'(r_+)\tau}{2}\Bigr)^2
+dR^2  \Bigr]+r_+^2 d \Omega_2^2.
\end{eqnarray}
%
We can see easily that the metric will be regular if we assume that the
$\tau$ direction is periodic with period $4\pi/U'(r_+)$.

In the case of $U(r)=1-r^2_0/r^2$ with $\lambda=\Lambda=0$, the exact five
dimensional solution for time-symmetric initial data ($K_{\mu\nu}=0$) is
%
\begin{eqnarray}
ds_5^2=-r^2dt^2+U(r)d\tau^2+U^{-1}(r)dr^2+r^2{\cosh}^2 t \, d\Omega_2^2.
\end{eqnarray}
%
This is the Witten-bubble spacetime\cite{KK1}. Another example of initial
data for a Kaluza-Klein bubble spacetime was given in Ref. \cite{KK2} and its
classical time evolution has been investigated in Ref. \cite{Ted,Shinkai}.

\if\answ\twocol
\begin{figure}[tbp]
\setlength{\unitlength}{1in}
\begin{picture}(3.0,7.5)
\put(0.0,5.25){\epsfxsize=3.0in \epsfysize=1.78in
\epsffile{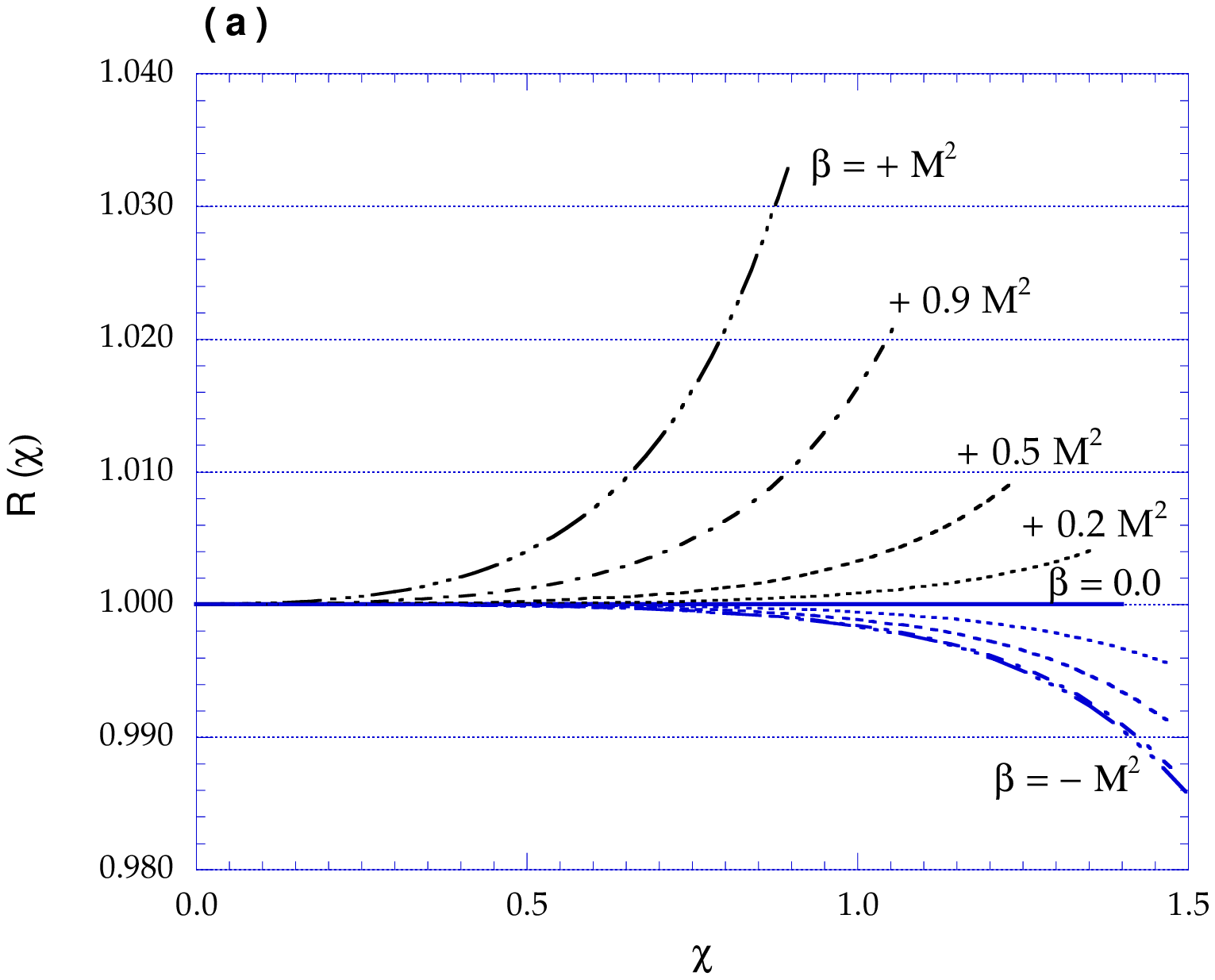} }
\put(0.0,2.75){\epsfxsize=3.0in \epsfysize=1.78in
\epsffile{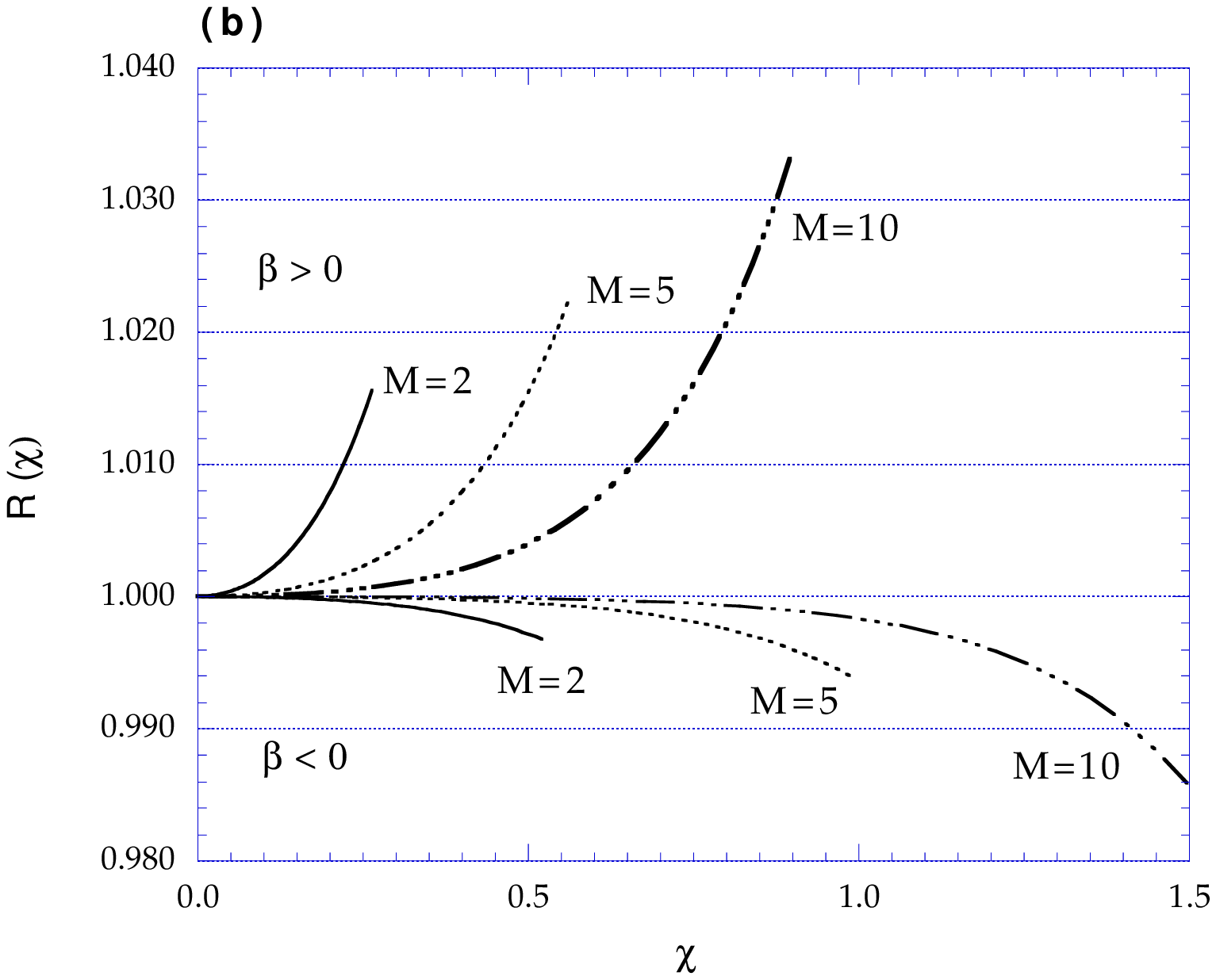} }
\put(0.0,0.25){\epsfxsize=3.0in \epsfysize=1.78in
\epsffile{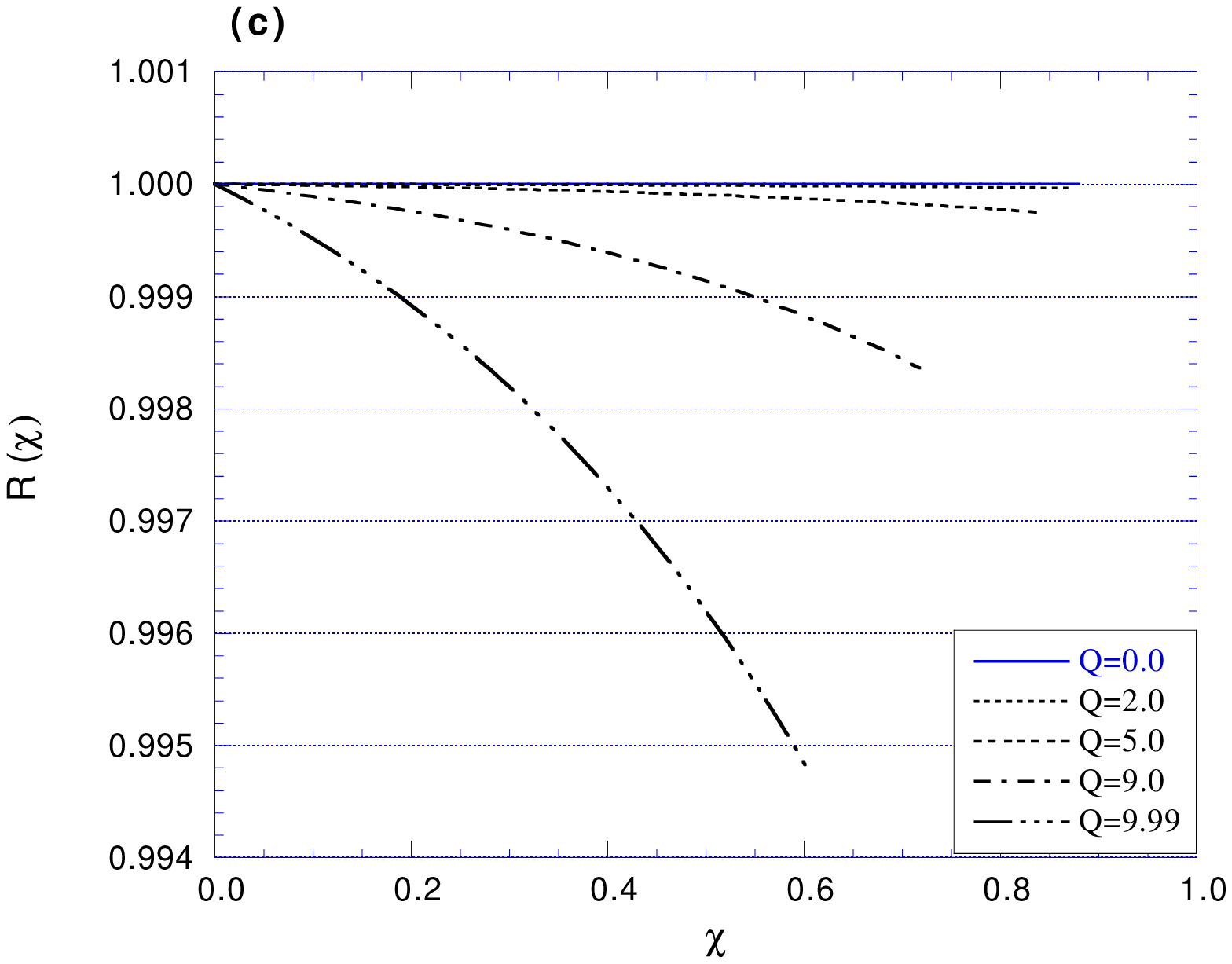} }
\end{picture}
\fi
\if\answ\prepri
\begin{figure}[p]
\setlength{\unitlength}{1in}
\begin{picture}(7.0,7.0)
\put(0.0,3.5){\epsfxsize=3.4in \epsfysize=2.1in \epsffile{fig1a-Qz.eps} }
\put(3.5,3.5){\epsfxsize=3.4in \epsfysize=2.1in \epsffile{fig1b-QzM.eps} }
\put(0.0,0.25){\epsfxsize=3.4in \epsfysize=2.1in \epsffile{fig1c-Bz.eps} }
\end{picture}
\fi

\caption[fig1]{
Ratio of the physical size of apparent horizon to size of
black string apparent horizon,
$R(\chi)$ [cf. eq. (\ref{AHstringratio})],  plotted as
a function of $\chi$.
Fig.(a) is for model (i).
Lines are of $\beta=0, \pm 0.2 M^2, \pm 0.5 M^2, \pm 0.9M^2$ and $\pm
M^2$,
where $M=10.0$.
We see that the qualitative behaviour of $R(\chi)$
depends on the sign of $\beta$.
Fig.(b) is for the extremal case of model (i), $\beta= M^2$
with different values of $M$. Results are also plotted for $\beta = -M^2$.
Fig.(c) is for model (ii) for which $R(\chi)$ is monotonically
decreasing.
}
\label{fig1}
\end{figure}

\if\answ\twocol
\begin{figure}[tbp]
\setlength{\unitlength}{1in}
   \begin{picture}(3.0,2.5)
   \put(0.0,0.25){\epsfxsize=3.0in \epsfysize=1.78in
                   \epsffile{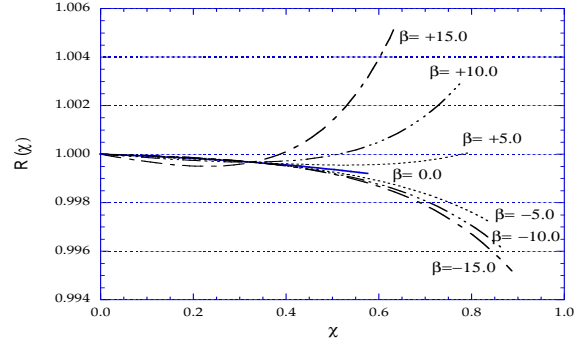} }
   \end{picture}
\fi
\if\answ\prepri
\begin{figure}[p]
\setlength{\unitlength}{1in}
     \begin{picture}(7.0,4.0)
     \put(1.0,0.25){\epsfxsize=4.0in \epsfysize=2.38in
                    \epsffile{fig2-Q3.eps}}
     \end{picture}
\fi

\caption[fig2]{
Ratio of physical size of apparent horizon to size of black string
apparent horizon,
$R(\chi)$ [cf. eq. (\ref{AHstringratio})], for nonzero $Q$ and $\beta$.
We have set $M=5$, $Q=3$ and $\beta=0, \pm 5, \pm 10, \pm 15$
for this plot.
The main features are a combination of plots in Fig.\ref{fig1}.
}
\label{fig2}
\end{figure}

\if\answ\twocol
\begin{figure}[tbp]
\setlength{\unitlength}{1in}
   \begin{picture}(3.0,2.5)
   \put(0.0,0.25){\epsfxsize=3.0in \epsfysize=1.78in
                   \epsffile{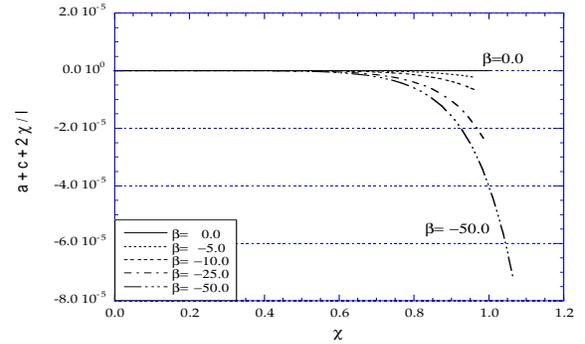} }
   \end{picture}
\fi
\if\answ\prepri
\begin{figure}[p]
\setlength{\unitlength}{1in}
     \begin{picture}(7.0,4.0)
     \put(1.0,0.25){\epsfxsize=4.0in \epsfysize=2.38in
                    \epsffile{fig3-Qz-expAC2X.eps}}
     \end{picture}
\fi

\caption[fig3]{
The quantity $a+c+2\chi/\ell$ at $r=r_+$ plotted
for $M=5, Q=0$ and $\beta=0, -5, -10, -25$ and $-50$.
}
\label{fig3}
\end{figure}

\if\answ\twocol
\begin{figure}[tbp]
\setlength{\unitlength}{1in}
   \begin{picture}(3.5,8.0)
   \put(-0.2,5.2){\epsfxsize=3.25in \epsfysize=3.25in
                   \epsffile{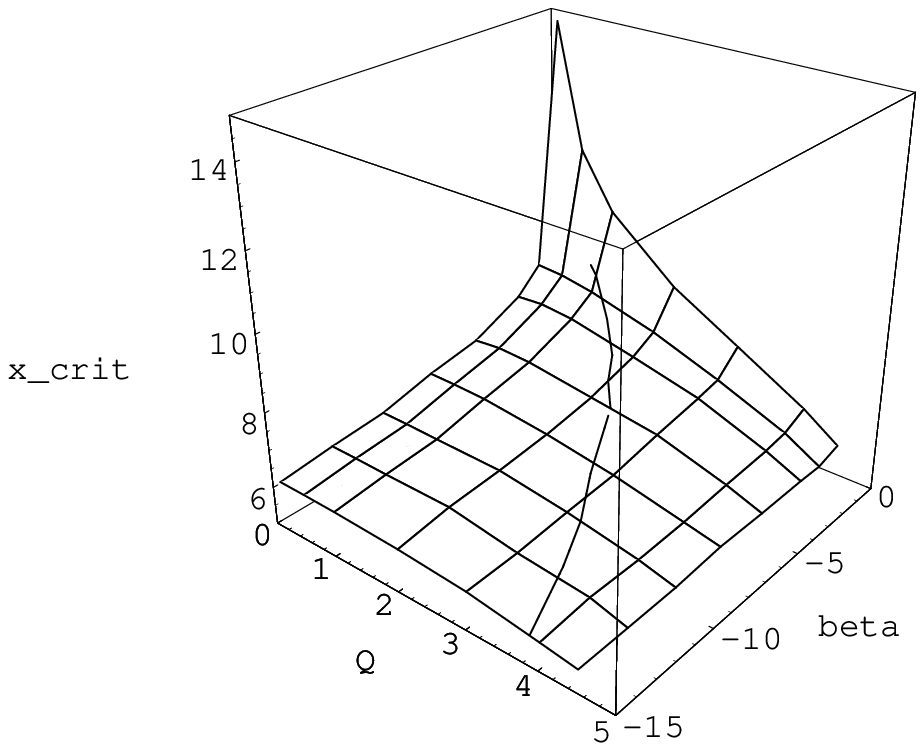} }
   \put(0.0,2.7){\epsfxsize=2.75in \epsfysize=2.75in
                   \epsffile{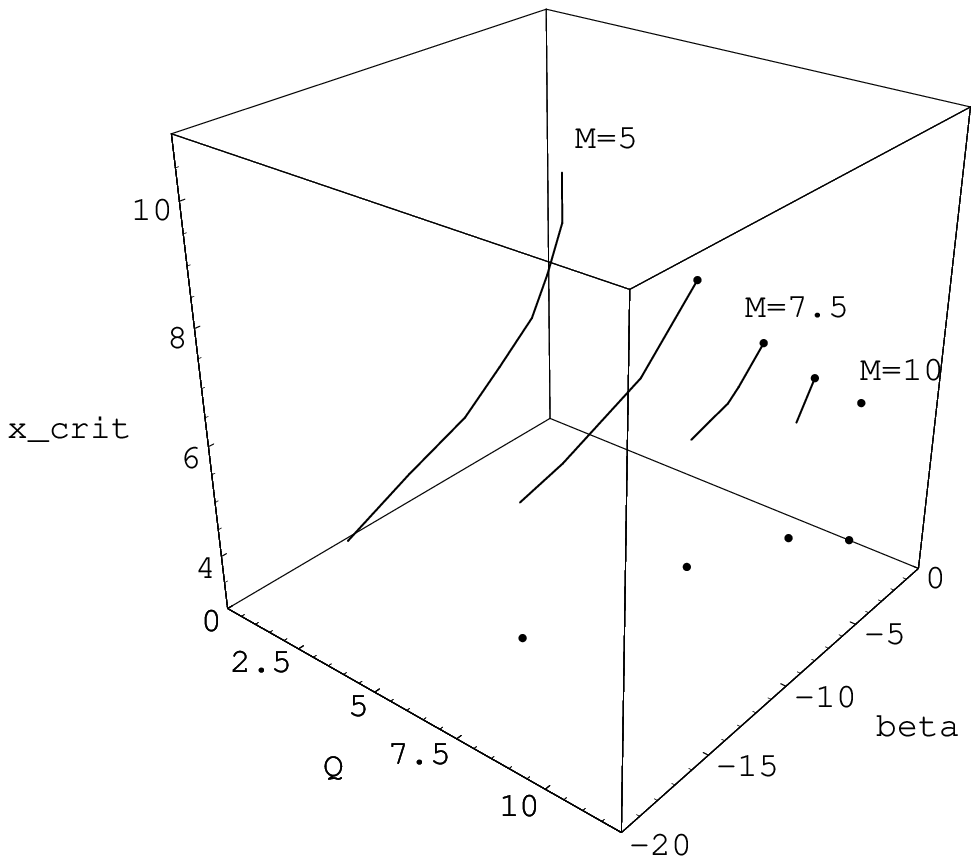} }
   \put(0.0,0.0){\epsfxsize=2.75in \epsfysize=2.75in
                   \epsffile{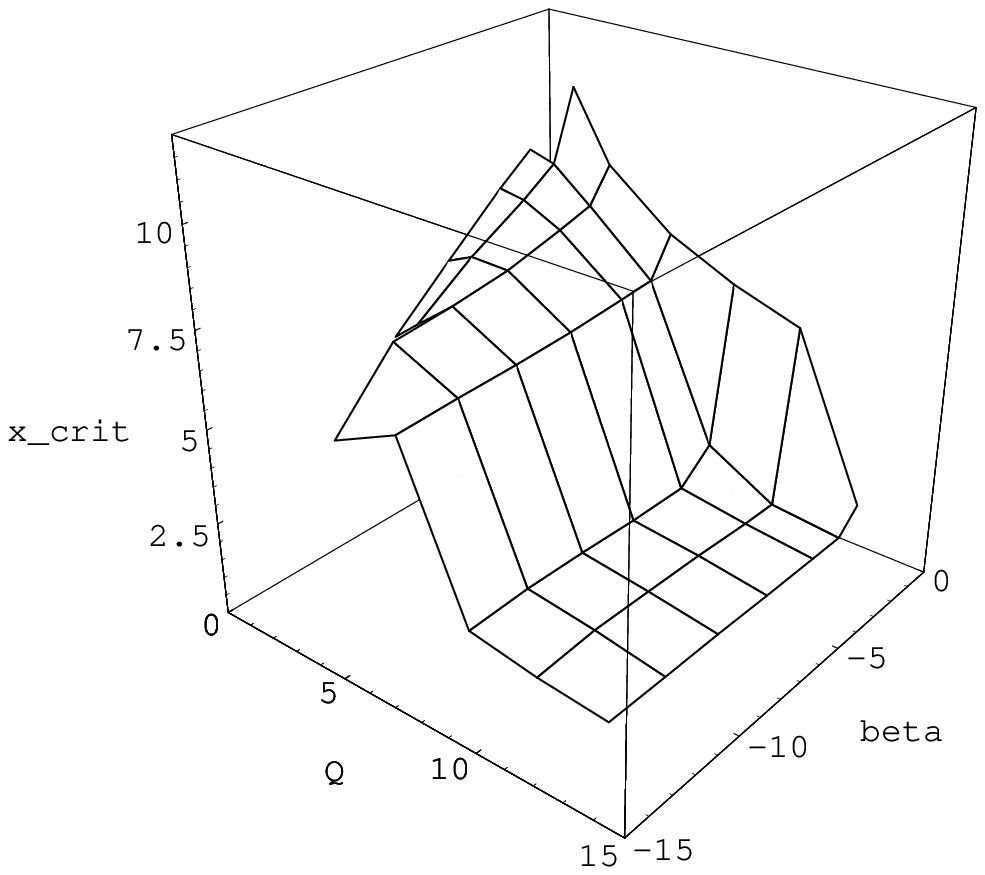}}
   \end{picture}
\fi
\if\answ\prepri
\begin{figure}[p]
\setlength{\unitlength}{1in}
      \begin{picture}(7.0,6.0)
     \put(0.25,3.0){\epsfxsize=3.5in \epsfysize=3.5in
                     \epsffile{fig4a.eps}}
     \put(3.75,3.25){\epsfxsize=3.0in \epsfysize=3.0in
                     \epsffile{fig4b.eps}}
     \put(0.5,0.25){\epsfxsize=3.0in \epsfysize=3.0in
                     \epsffile{fig4c.eps}}
     \end{picture}
\fi

\caption[fig4]{
Critical value $\chi_{crit}$ [eq.(\ref{eq-Xcrit})] plotted for
non-zero $Q$ and $\beta (\leq 0)$ black holes.
(a) [above left] We have set $M=5$ for this plot.
Note that for the uncharged case,  $\chi_{crit}=\infty$.
The solid curve is for the special cases with $r_+=10.0$.
(b) [above right] Critical value $\chi_{crit}$
for combinations of parameters ($Q,\beta$) which produce a black hole
with $r_+=10.0$.  $M=5.0, 6.25, 7.5, 8.75$ and  $10.0$ are chosen for
these plots. The black dots denote the ends of the lines at $\beta=0$
and the other ends projected onto the  $\chi_{crit}=3$ plane.
(c) [below] The same plot as for (b), but $M$ is specified so as to fix
$r_+=10$
for given ($Q,\beta$).
}

\label{fig4}
\end{figure}



\begin{thebibliography}{66}
\bibitem{add}
N. Arkani-Hamed, S. Dimopoulos and G. Dvali, Phys. Lett. {\bf B429},
263 (1998).
\bibitem{aadd}
I. Antoniadis, N. Arkani-Hamed, S. Dimopoulos and G. Dvali,
Phys. Lett. {\bf B436}, 257 (1998).
\bibitem{RS1}
L.~Randall and R.~Sundrum, Phys. Rev. Lett. {\bf 83}, 3370(1999).
\bibitem{RS2}
L.~Randall and R.~Sundrum, Phys. Rev. Lett. {\bf 83}, 4690(1999).
\bibitem{old}
V. A. Rubakov and M. E. Shaposhinikov, Phys. Lett. {\bf 152B}, 136(1983);\\
M. Visser, Phys. Lett. {\bf 159B}, 22(1985);\\
M. Gogberashvili, Mod. Phys. Lett. {\bf A14}, 2025(1999).
\bibitem{Andrew}
A.~Chamblin, S.~W.~Hawking and H.~S.~Reall,
Phys. Rev. {\bf D61}, 065007 (2000). 
\bibitem{Ruth}
R.~Gregory and R.~Laflamme, Phys. Rev. Lett. {\bf 70}, 2837(1993);\\
R.~Gregory, hep-th/0004101.
\bibitem{csaba}
A. Chamblin, C. Csaki, J. Erlich and T.J. Hollowood, Phys.Rev. {\bf D62},
044012 (2000).
\bibitem{Gary}
R.~Emparan, G.~T.~Horowitz, and R.~C.~Myers, JHEP {\bf 0001}, 007 (2000).
\bibitem{Dad}
N. Dadhich, R. Maartens, P. Papadopoulos and V. Rezania, hep-th/0003061.
\bibitem{Tess}
T.~Shiromizu, K.~Maeda and M.~Sasaki, Phys. Rev. {\bf D62}, 024012 (2000).
\bibitem{Gubser}
S.~S.~Gubser, hep-th/9912001.
\bibitem{kaloper}
N. Kaloper, E. Silverstein and L. Susskind, hep-th/0006192.
\bibitem{lu}
H. Lu and C.N. Pope, hep-th/0008050.
\bibitem{oda}
I. Oda, hep-th/0008055.
\bibitem{Shibata}
T. Shiromizu and M. Shibata, hep-th/0007203.
\bibitem{Israel}
W.~Israel, Nuovo. Cimento. {\bf 44B}, 1(1966); erratum: {\bf 48B}, 463
(1967).
\bibitem{Tama}
J.~Garriga and T.~Tanaka, Phys. Rev. Lett. {\bf 84}, 2778 (2000).
\bibitem{Misao}
M.~Sasaki, T.~Shiromizu and K.~Maeda, Phys. Rev. {\bf D62}, 024008 (2000).
\bibitem{Lisa}
S.~B.~Giddings, E.~Katz, and L.~Randall, JHEP {\bf 0003}, 023 (2000).
\bibitem{KK1}
E. Witten, Nucl. Phys. {\bf B195}, 481 (1982).
\bibitem{KK2}
D. Brill and G. T. Horowitz, Phys. Lett. {\bf 262}, 437 (1991).
\bibitem{Ted}
S.~Corley and T.~Jacobson, Phys. Rev. {\bf D49}, 6261(1994).
\bibitem{Shinkai}
H.~Shinkai and T.~Shiromizu, Phys. Rev. {\bf D62}, 024010 (2000).
\bibitem{TeukolskyCrankNicholson}
{\it e.g.,} S.A. Teukolsky, Phys. Rev. {\bf D61}, 087501 (2000).
\bibitem{HawkingEllis}
S. W. Hawking and G. F. R. Ellis, {\it The large scale structure of
space-time}, (Cambridge Univ. Press, 1973);\\
R. M. Wald, {\it General Relativity}, (Univ. of Chicago Press, 1984).

\end{thebibliography}
\end{document}